\documentclass[acmsmall,screen=true]{acmart}

\usepackage[ruled,vlined,lined,commentsnumbered]{algorithm2e}
\usepackage{amsmath,amsfonts}
\usepackage{array}
\usepackage{booktabs}
\usepackage[skip=1pt,labelfont=bf]{caption}
 \usepackage{calligra}
\usepackage{color, colortbl}
\usepackage{courier}
\usepackage{csvsimple}
\usepackage{enumitem}
\usepackage{fancybox}
\usepackage{fontenc}
\usepackage{graphicx}
\usepackage{listings}
\usepackage{longtable}
\usepackage{lscape}
\usepackage{makecell}
\usepackage{marvosym}
\usepackage{moreverb}
\usepackage{multicol}
\usepackage{multirow}
\usepackage{pifont}
\usepackage{rotating}
\usepackage{setspace}
\usepackage{caption}
\usepackage{subfigure}
\usepackage[most]{tcolorbox}
\usepackage{threeparttable}
\usepackage{tikz}
\usepackage[normalem]{ulem}
\usepackage{url}
\usepackage{soul}
\usepackage{wasysym}
\usepackage{svg}
\usepackage{textcomp}
\usepackage{xcolor}
\usepackage{wrapfig}
\usepackage{pdflscape}
\usepackage{hyperref}

\usepackage{amsmath}
\usepackage{array}
\usepackage{balance}
\usepackage{microtype}
\usepackage{afterpage}
\newcolumntype{L}[1]{>{\raggedright\let\newline\\\arraybackslash\hspace{0pt}}m{#1}}
\newcolumntype{C}[1]{>{\centering\let\newline\\\arraybackslash\hspace{0pt}}m{#1}}
\newcolumntype{R}[1]{>{\raggedleft\let\newline\\\arraybackslash\hspace{0pt}}m{#1}}

\newcommand{\intuition}[1]{
\begin{tcolorbox}[colback=white,boxrule=1pt,top=0pt,bottom=0pt,left=1pt,right=2pt,top=2pt,bottom=2pt]
\em #1
\end{tcolorbox}
}

\AtBeginDocument{%
  \providecommand\BibTeX{{%
    \normalfont B\kern-0.5em{\scshape i\kern-0.25em b}\kern-0.8em\TeX}}}

\begin{document}

\title{Improving the Ability of Pre-trained Language Model by Imparting Large Language Model's Experience}

\author{Xin Yin}
\affiliation{%
  \institution{The State Key Laboratory of Blockchain and Data Security, Zhejiang University}
  \city{Hangzhou}
  \country{China}
  }
\email{xyin@zju.edu.cn}

\author{Chao Ni}
\authornote{Chao Ni the corresponding author.\\
He is also with Hangzhou High-Tech Zone (Binjiang) Institute of Blockchain and Data Security.}
\affiliation{%
  \institution{The State Key Laboratory of Blockchain and Data Security, Zhejiang University}
  \city{Hangzhou}
  \country{China}
  }
\email{chaoni@zju.edu.cn}

\author{Xiaodan Xu}
\affiliation{%
  \institution{The State Key Laboratory of Blockchain and Data Security, Zhejiang University}
  \city{Hangzhou}
  \country{China}
}
\email{xiaodanxu@zju.edu.cn}

\author{Xinrui Li}
\affiliation{%
  \institution{The State Key Laboratory of Blockchain and Data Security, Zhejiang University}
  \city{Hangzhou}
  \country{China}
  }
\email{lixinrui@zju.edu.cn}

\author{Xiaohu Yang}
\affiliation{%
  \institution{The State Key Laboratory of Blockchain and Data Security, Zhejiang University}
  \city{Hangzhou}
  \country{China}
}
\email{yangxh@zju.edu.cn}

\begin{abstract}

Large Language Models (LLMs) and pre-trained Language Models (LMs) have achieved impressive success on many software engineering tasks (e.g., code completion and code generation).
By leveraging huge existing code corpora (e.g., GitHub), these models can understand the patterns in source code and use these patterns to predict code properties.
However, LLMs under few-shot learning perform poorly on non-generative tasks (e.g., fault localization and vulnerability localization), and fine-tuning LLMs is time-consuming and costly for end users and small organizations.
Furthermore, the performance of fine-tuning LMs for non-generative tasks is impressive, yet it heavily depends on the amount and quality of data.
As a result, the current lack of data and the high cost of collecting it in real-world scenarios further limit the applicability of LMs.
In this paper, we leverage the powerful generation capabilities of LLMs to enhance pre-trained LMs. 
Specifically, we use LLMs to generate domain-specific data, thereby improving the performance of pre-trained LMs on the target tasks.
We conduct experiments by combining different LLMs in our generation phase and introducing various LMs to learn from the LLM-generated data. 
Then, we compare the performance of these LMs before and after learning the data.
We find that LLM-generated data significantly enhances the performance of LMs. 
The improvement can reach up to 58.36\% for fault localization and up to 6.09\% for clone detection.

\end{abstract}



\maketitle

\section{Introduction}

Large Language Models (LLMs)~\cite{roziere2023code,deepseek-coder,zheng2024opencodeinterpreter,luo2023wizardcoder,wei2023magicoder,phi,meta2024llama,jiang2023mistral} have been widely adopted due to advances in Natural Language Processing (NLP). 
These advances allow LLMs to be trained with billions of parameters and samples, resulting in significant performance improvements on various tasks.
LLMs can easily be used for code-related tasks by being fine-tuned~\cite{zhang2024empirical,yin2024multitask} or prompted~\cite{xia2023keep,xia2023automated,yin2024multitask,yin2024thinkrepair} since they are trained to be general and can capture knowledge from various domains.
LLMs perform better as their computational budget increases~\cite{kaplan2020scaling}. 
For instance, their performance on program synthesis benchmarks improves linearly with the number of model parameters~\cite{xu2022systematic}.
However, many existing works in software engineering, such as fault localization and vulnerability detection, either train small models from scratch~\cite{li2019deepfl,li2021fault} or fine-tune modest-sized models~\cite{meng2022improving,ni2023distinguishing}. 
They often overlook the potential benefits of fine-tuning state-of-the-art LLMs.
This oversight occurs because fine-tuning LLMs is time-consuming and computationally expensive for end users and small organizations.

Recently, pre-trained Language Models (LMs) have shown remarkable success in various software engineering tasks~\cite{feng2020codebert,guo2020graphcodebert,guo2022unixcoder,ahmad2021unified,wang2021codet5,wang2023codet5+}.
By leveraging huge existing code corpora (e.g., GitHub), these models can understand the patterns in source code and use these patterns to predict code properties.
These models are small but powerful, usually having between 110M and 220M parameters.
Therefore, they require less time and computational resources than LLMs to fine-tune for various software engineering tasks, and their effectiveness has been widely demonstrated~\cite{ni2023distinguishing,xia2022less,fu2022linevul,hin2022linevd}. 
However, their performance significantly relies heavily on the amount and quality of data.
The current scarcity of data, coupled with the high cost of data collection in real-world scenarios, presents a major challenge to the practical application of these models.
Traditional generation tools can be used to generate data, but different tools are required to generate data specific to different tasks. 
There is no unified tool for generating data across all tasks, and many tools are based on specific patterns for data generation. 
For example, mutation testing tools can be used to generate faults, but they typically rely on predefined patterns to create mutations. 
These patterns are manually designed to simulate common programming errors or code changes, yet they may not cover all possible code variations.

Benefiting from pre-training on large-scale code corpora, LLMs possess the ability to generate large-scale and diverse code (e.g., buggy function and code clone).
To address the challenge of data scarcity, we leverage the powerful generation capabilities of LLMs to enhance pre-trained LMs.
Specifically, we use LLMs to generate domain-specific data, thereby improving the performance of pre-trained LMs on the target tasks.

We evaluate the LLMs' capability to improve the LMs on two non-generative tasks (i.e., fault localization and clone detection) by using eight popular LLMs to generate faults and clones from Defects4J~\cite{just2014defects4j} and HumanEval-X~\cite{zheng2023codegeex} datasets, which can be divided into two categories: five code LLMs (i.e., Magicoder~\cite{wei2023magicoder}, DeepSeek-Coder~\cite{deepseek-coder}, OpenCodeInterpreter~\cite{zheng2024opencodeinterpreter}, CodeLlama~\cite{roziere2023code}, and WizardCoder~\cite{luo2023wizardcoder}) and three general LLMs (i.e., Llama 3~\cite{meta2024llama}, Mistral~\cite{jiang2023mistral}, and Phi-2~\cite{phi}).
We then design selection strategies to identify high-quality data and introduce various LMs to learn from the LLM-generated data. 
The LMs we utilize have fewer than 220M parameters and can be categorized into two types: encoder-only LMs (i.e., BERT~\cite{devlin2018bert}, RoBERTa~\cite{liu2019roberta}, CodeBERT~\cite{feng2020codebert}, GraphCodeBERT~\cite{guo2020graphcodebert}, and UniXcoder~\cite{guo2022unixcoder}) and encoder-decoder LMs (i.e., PLBART~\cite{ahmad2021unified}, CodeT5~\cite{wang2021codet5}, and CodeT5+~\cite{wang2023codet5+}).

We compare the performance of these LMs before and after learning the LLM-generated data.
We find that LLM-generated data significantly enhances the performance of LMs. 
The improvement can reach up to 58.36\% for fault localization and up to 6.09\% for clone detection. 
For example, in fault localization, by adding 30\% of the generated data to the training set, GraphCodeBERT’s F1-score increases from 0.384 to 0.550, Recall increases from 0.353 to 0.522, Precision increases from 0.421 to 0.582, and Accuracy increases from 0.889 to 0.901.
Moreover, after learning from the LLM-generated data, LMs show an improvement of 0.74\%$\sim$6.09\% on the CodeXGLEU-POJ104~\cite{lu2021codexglue} dataset. 
Similarly, on the CodeNet-Java250~\cite{puri2021codenet} dataset, LMs exhibited an increase of 0.13\%$\sim$3.23\%. 
Our study highlights that using LLMs to generate data for LMs can improve performance by a large margin.

To summarize, the main contributions of this paper are:

\begin{itemize}[leftmargin=*]

\item We generate additional data for fault localization and code clone detection using LLMs and employ selection strategies to ensure data quality.

\item We improve the capabilities of LMs on non-generative tasks using LLM-generated data.
Our evaluation demonstrates that the LMs' effectiveness improves significantly after learning from the generated data.

\item We conduct comprehensive experiments on eight LLMs and eight LMs using widely studied datasets~\cite{just2014defects4j,madeiral2019bears,saha2018bugs,zheng2023codegeex,lu2021codexglue,puri2021codenet} to explore their effectiveness.

\end{itemize}
\label{sec:introduction}

\section{Related Work}
\label{sec:related_work}

\subsection{Language Model}
Language Models can perform tasks such as clone detection and code generation.
We discuss such models by summarizing them into pre-trained Language Models and Large Language Models.

\noindent
\textbf{Pre-trained Language Models.}
Pre-trained Language Models (LMs) are usually trained on a large volume of data and can be classified into two types of architectures: encoder-only and encoder-decoder. 
Encoder-only (e.g., BERT~\cite{devlin2018bert} and  CodeBERT~\cite{feng2020codebert}) and encoder-decoder (e.g., PLBART~\cite{ahmad2021unified} and CodeT5~\cite{wang2021codet5}) models are trained using Masked Language Modeling (MLM) or Masked Span Prediction (MSP) objective, respectively, where a small portion (e.g., 15\%) of the tokens are replaced with either masked tokens or masked span tokens, models are trained to recover the masked tokens.
Apart from tasks in NL (e.g., cloze test and question answering), they are also widely used in code-related tasks (e.g., fault localization and clone detection).
These models are small but powerful, usually having between 110M and 220M parameters.

\noindent
\textbf{Large Language Models.}
Large Language Models (LLMs)~\cite{roziere2023code,deepseek-coder,zheng2024opencodeinterpreter,luo2023wizardcoder,wei2023magicoder,phi,meta2024llama,jiang2023mistral} have been widely adopted since the advances in Natural Language Processing which enable LLMs to be well-trained with both billions of parameters and billions of training samples, which consequently brings a large performance improvement on tasks adopted by LLMs~\cite{yin2024you,bang2023multitask,ouyang2022training}.
These models can be easily used for a downstream task by being fine-tuned~\cite{zhang2024empirical,yin2024multitask} or being prompted~\cite{xia2023keep,xia2023automated,yin2024rectifier,yin2024thinkrepair,ni2024learning} since they are trained to be general and they can capture different knowledge from various domains.
Fine-tuning is used to update model parameters for a particular downstream task by iterating the model on a specific dataset while prompting can be directly used by providing natural language descriptions or a few examples of the downstream task.
Compared to prompting, fine-tuning is expensive since it requires additional model training and has limited usage scenarios, especially in cases where sufficient training datasets are unavailable.

\subsection{Machine Learning Fault Localization}
Machine Learning Fault Localization (MLFL) techniques use program analysis to understand code behavior.
They've used various data types like test coverage metrics~\cite{briand2007using,zhang2017deep}, co-changing method declarations~\cite{li2022fault}, and structural information from the code, such as the abstract syntax tree (AST)~\cite{li2021fault}. 
Recent approaches, like GRACE~\cite{lou2021boosting} and FixLocator~\cite{li2022fault}, encode both AST and test coverage into graph representations. 
These methods prioritize faulty methods by preserving all topological dependencies with graph neural networks.
DeepRL4FL~\cite{li2021fault} uses a convolution neural network to analyze code coverage matrices.
DeepFL~\cite{li2019deepfl} and TRANSFER-FL~\cite{meng2022improving} integrate features based on semantics, spectrum, and mutations using a multi-layer perceptron model. 
LLMAO~\cite{yang2024large}, unlike previous techniques, doesn't need test code or an AST parser. 
It includes a text tokenizer and embedding layer and leverages attention mechanisms and bidirectional adapter layers on pre-trained left-to-right LLMs directly on source code.

\subsection{Code Clone Detection}
Code clone detection aims to extract similar pairs of code snippets from large code bases.
Several novel neural models~\cite{wei2017supervised, zhang2019novel,wang2020detecting} have been proposed by the community for this purpose, leveraging abstract syntax tree or data flow information obtained from compilers to comprehend code functionality. 
Pre-trained LMs have achieved impressive success on many software engineering tasks~\cite{guo2022unixcoder,wang2021codet5,wang2023codet5+}, including code clone detection, through fine-tuning.
During the fine-tuning stage, code snippets undergo encoding into low-dimensional dense vectors. 
In this process, the similarity between vectors in the latent space is amplified for clone pairs, while it is diminished for non-clone pairs.

\section{Approach}
\label{sec:approach}

\begin{figure}[htbp]
    \centering
    \includegraphics[width=.9\linewidth]{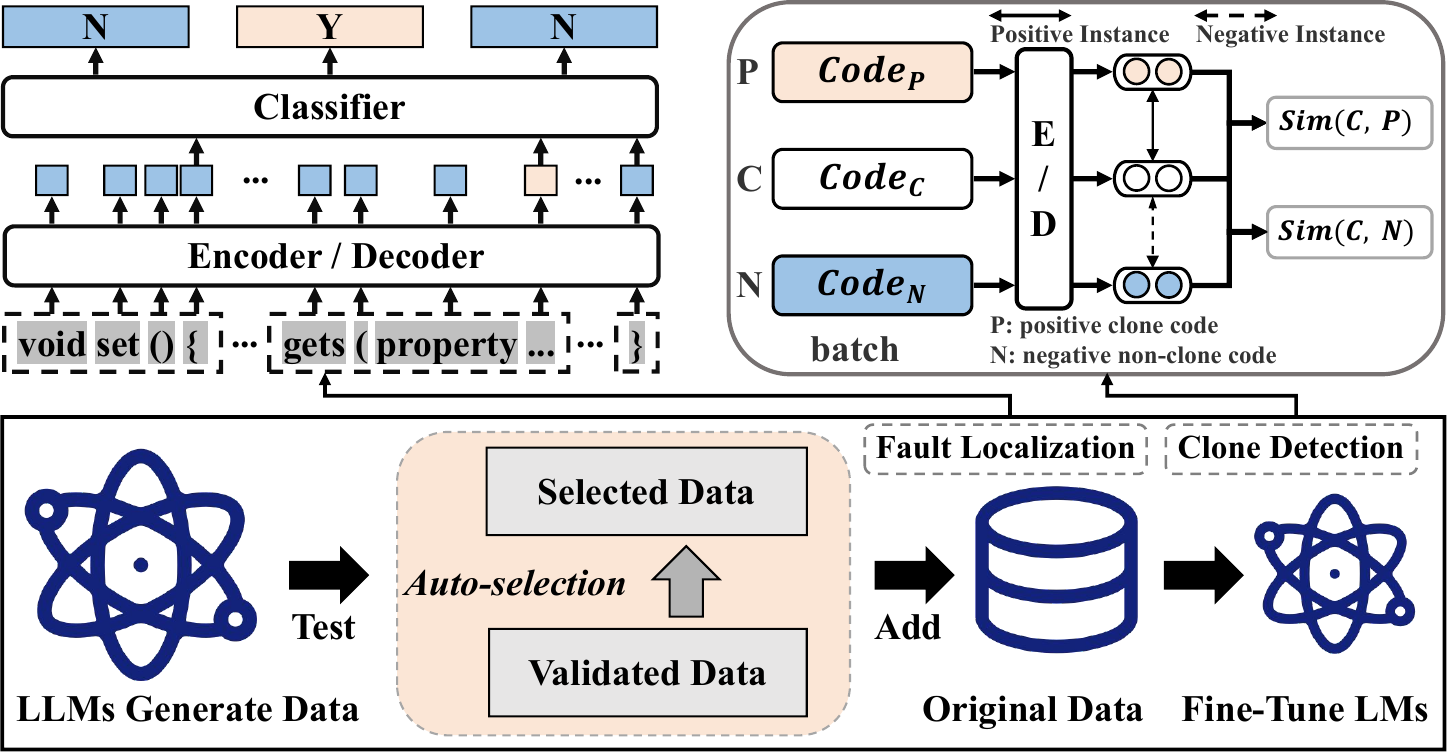}
    \caption{Fine-tune LMs to learn from LLM-generated data}
   \label{fig:overview}
\end{figure}

In this section, we begin by discussing the models selected for generation and evaluation. 
Then, we demonstrate practical methods for using LLMs to generate data, along with selection strategies to ensure data quality.
Finally, we describe techniques for fine-tuning pre-trained LMs in fault localization and clone detection tasks.
The overview of our approach is shown in Fig.~\ref{fig:overview}.

\subsection{Studied Language Models}

We describe the various pre-trained LMs we use for evaluation.
As shown in Table~\ref{tab:studied_models}, these models have fewer than 220 million parameters and can be categorized into two categories: encoder-only LMs and encoder-decoder LMs.
Encoder-only LMs (i.e., BERT~\cite{devlin2018bert}, RoBERTa~\cite{liu2019roberta}, CodeBERT~\cite{feng2020codebert}, GraphCodeBERT~\cite{guo2020graphcodebert}, and UniXcoder~\cite{guo2022unixcoder}) contain only the encoder component of a Transformer. 
They are designed for learning data representations and trained using the Masked Language Modeling (MLM) objective. 
Encoder-decoder LMs (i.e., PLBART~\cite{ahmad2021unified}, CodeT5~\cite{wang2021codet5}, and CodeT5+~\cite{wang2023codet5+}) have been proposed for sequence-to-sequence tasks.
They are trained to recover the correct output sequence given the original input, often through span prediction tasks where random spans are replaced with artificial tokens. 
All these models can potentially be used for classification in our tasks, so we evaluate these state-of-the-art LMs.

\begin{table}[htbp]
  \centering
  \caption{Overview of the studied models}
  \resizebox{\linewidth}{!}
  {
    \begin{threeparttable}
    \begin{tabular}{ccc|ccc}
    \toprule
    \textbf{Studied LMs} & \textbf{\# Para.} & \textbf{Model Type} & \textbf{Studied LLMs} & \textbf{\# Para.} & \textbf{Model Type} \\
    \midrule
    BERT & 110M & Encoder-only LM & Magicoder & 6.7B & Code LLM \\
    RoBERTa & 125M & Encoder-only LM & DeepSeek-Coder & 6.7B & Code LLM \\
    CodeBERT & 125M & Encoder-only LM & OpenCodeInterpreter & 6.7B & Code LLM \\
    GraphCodeBERT & 125M & Encoder-only LM & CodeLlama & 7B & Code LLM \\
    UniXcoder & 125M & Encoder-only LM & WizardCoder & 7B & Code LLM \\
    PLBART & 140M & Encoder-decoder LM & Phi-2 & 2.7B & General LLM \\
    CodeT5 & 220M & Encoder-decoder LM & Mistral & 7B & General LLM \\
    CodeT5+ & 220M & Encoder-decoder LM & Llama 3 & 8B & General LLM \\
    \bottomrule
    \end{tabular}%
    $^\ast$For UniXcoder, we use encoder-only mode.
    \end{threeparttable}
  }
  \label{tab:studied_models}%
\end{table}%

\subsection{Studied Large Language Models}
We describe the various open-source LLMs we employ for evaluation.
General LLMs are trained on textual data, encompassing both natural language and code, and are versatile for numerous tasks. 
Conversely, code LLMs are tailored specifically for automating code-related tasks.
Among the code LLMs, we select the five models released recently (in 2024), namely Magicoder~\cite{wei2023magicoder}, DeepSeek-Coder~\cite{deepseek-coder}, OpenCodeInterpreter~\cite{zheng2024opencodeinterpreter}, CodeLlama~\cite{roziere2023code}, and WizardCoder~\cite{luo2023wizardcoder}. 
For the general LLMs, we choose the top three models: Llama 3~\cite{meta2024llama}, Mistral~\cite{jiang2023mistral}, and Phi-2~\cite{phi}.
A summary of the characteristics of these selected LLMs is presented in Table~\ref{tab:studied_models}.

\subsection{LLM-based Data Generation}
In our study, we use two non-generative tasks for evaluation: fault localization and clone detection.
We now describe how to generate data for these tasks.

\subsubsection{Fault Generation}
Since we evaluate function-level fault localization, we use LLMs to autoregressively inject faults into the original non-buggy functions. 
However, since LLMs are not specifically pre-trained for this task, simply providing them with the non-buggy function will not suffice.
To facilitate the direct usage of LLMs for fault generation, we use a specific prompt to enable the LLMs to perform few-shot learning. 
This allows the LLMs to recognize the task and generate a buggy function by completing the input provided. 
We follow the prompt similar to those used in the artifacts, papers, or technical reports~\cite{shieh2023best,nijkamp2022codegen, li2023starcoder,roziere2023code,yin2024multitask}.

To help LLM perform downstream tasks effectively, a well-crafted prompt is crucial, a topic explored by various researchers~\cite{xia2023keep,white2023prompt,feng2023prompting}. 
Previous research~\cite{zhang2022automatic} suggests that including a few diverse examples can enhance LLM's generalization ability. 
To achieve this, we employ an advanced selection strategy (i.e., Semantic-based Selection) to select semantically similar examples.
This selection strategy adopts a pre-trained model (i.e., UniXcoder, which effectively comprehends code semantic information~\cite{guo2022unixcoder}) to embed all the functions and then uses the K-means algorithm~\cite{macqueen1967some} for clustering. 
We cluster the examples (i.e., pairs of non-buggy and buggy functions) from the training set of Defects4J based on their semantic similarity to ensure the selection of distinct examples~\cite{zhang2022automatic,li2023mot}. 
Considering the limitation of LLM's conversation windows, we adopt a two-shot experimental setup: we cluster all examples into two clusters and then select the most semantically similar example from each cluster based on cosine similarity.

\begin{figure}[htbp]
    \centering
    \includegraphics[width=.8\linewidth]{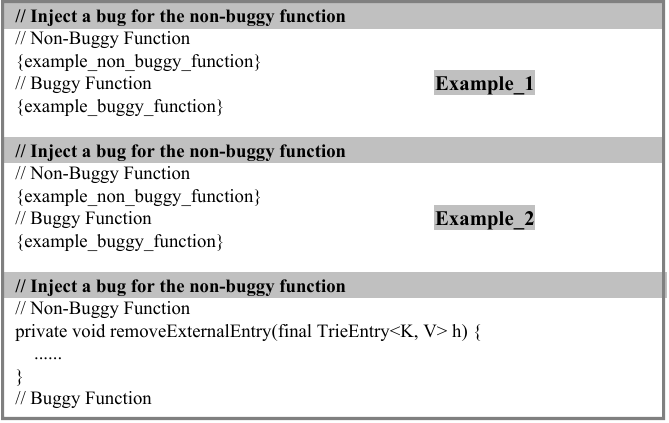}
    \caption{An example of prompt for fault generation}
    \label{fig:prompt}
\end{figure}

Fig.~\ref{fig:prompt} illustrates the prompt, consisting of two fault generation examples designed to showcase the task and the desired output format.
Each example contains four elements: (1) task description, (2) non-buggy function, (3) indicator, and (4) buggy function, while input contains only the first three elements.

\begin{itemize}[leftmargin=*]

\item \textbf{Task Description}. 
We provide LLM with the description constructed as \textit{``// Inject a bug for the non-buggy function''}. 
The task descriptions used in the fault generation task vary based on the source programming language we employ.

\item \textbf{Non-Buggy Function}. 
We provide LLM with the non-buggy function. 
We also prefix with \textit{``// Non-Buggy Function''} to directly indicate LLM about the context of the function.

\item \textbf{Indicator}. 
We instruct LLM to think about the results.
We follow the best practice in previous works~\cite{xia2023automated,yin2024multitask} and adopt the similar indicator named \textit{``// Buggy Function''}.

\item \textbf{Buggy Function}. 
We demonstrate the expected output in the example, showing the buggy function produced after injecting faults into the non-buggy function.

\end{itemize}

Given a corpus of non-buggy functions, we use the decorated prompt to collect buggy functions by injecting faults into the non-buggy functions.

\subsubsection{Clone Generation}
Previous studies~\cite{luo2023wizardcoder,deepseek-coder,zheng2024opencodeinterpreter,wang2023codet5+,roziere2023code,wei2023magicoder} have assessed the ability of LLMs to generate code from natural language specifications in a zero-shot setting. 
Similarly, in clone generation, we can guide LLMs to generate similar code clones for the same natural language specification. 
Leveraging the official prompts from HumanEval-X~\cite{zheng2023codegeex}, we generate clones in both C++ and Java languages.

\subsection{Data Validation and Selection}
\label{sec:selection}
To effectively evaluate the data generated by LLMs, we need to verify and filter out invalid data, and then select high-quality data for evaluation.

\subsubsection{Data Validation.}
We compile and run test suites (originally supported by the studied datasets, cf. Section~\ref{sec:dataset}) to verify all candidate code snippets generated by LLM.
For fault generation, we retain only the buggy functions that fail to pass the test cases. 
For clone generation, we keep the generated functions that successfully pass the entire test suite.

\subsubsection{Data Selection.}
To select high-quality fault data, we analyze the Defects4J training set across three dimensions: (1) the average number of lines changed ($LC_{ave}$), determined by comparing file differences using the git diff command; (2) the average edit distance ($ED_{ave}$), calculated using Levenshtein edit distance between non-buggy and buggy functions; and (3) the average semantic similarity ($SS_{ave}$), obtained by inputting non-buggy and buggy functions into CodeBERT to retrieve vector representations and then computing cosine similarity.
We then compute $LC$, $ED$, and $SS$ for each generated data and calculate a score using the formula: $Score=|(LC-LC_{ave})/LC_{ave}|+|(ED-ED_{ave})/ED_{ave}|+|(SS-SS_{ave})/SS_{ave}|$. 
A lower score indicates closer alignment with the Detects4J training set's distribution.
Since LLMs generate multiple buggy functions for each non-buggy function, to ensure diversity, we select the buggy function with the lowest score from the generated data, resulting in a unique pair of non-buggy and buggy functions.
Finally, we select a subset or all of the data based on the $Score$, ranging from low to high (e.g., 10\%×Generated in Section~\ref{sec:rq3}).

For generated clone data, our goal is to maximize differences. 
We calculate the average edit distance among all generated code clones for each task in HumanEval-X. 
Additionally, we compute the average relative edit distance for each code with other code clones, selecting those with an average relative edit distance greater than or equal to the overall average edit distance.

\subsection{Model Fine-Tuning}

As illustrated in Fig.~\ref{fig:overview}, we demonstrate our process of selecting data from validated data and integrating it into the original dataset. 
Next, we fine-tune the LMs on two downstream tasks: fault localization and clone detection. 
We utilize eight LMs mentioned in Table~\ref{tab:studied_models} for our study, though other LMs like CuBERT~\cite{kanade2020learning} can be used interchangeably.

For fault localization, we follow previous works~\cite{yang2024large,zhang2024empirical} that use LMs to classify individual code lines as either buggy or non-buggy. 
For a token sequence $T=\{t_1, t_2,...,t_n\}$ of the function, the model’s encoder or decoder component, denoted as $M$, processes $T$ to yield a sequence of output vectors: $O=M(T)=\{o_1, o_2,..., o_L\}$, where $O$ represents the output tensor with dimensions $L \times H$, $L$ signifies the sequence length, and $H$ denotes the hidden dimension size.
During the process, the contextual information is captured by the self-attention or masked self-attention mechanisms in the encoder or decoder of LMs, where masked self-attention limits the sight to the preceding part of tokens. 
Each output vector $o_i$ that represents the last token of one line is subsequently associated with a label (i.e., 0 or 1). 
The optimization process employs the binary cross-entropy as the loss function.

Regarding clone detection, we follow prior works~\cite{guo2022unixcoder,wang2021codet5,wang2023codet5+} that directly use LMs to obtain embeddings of code snippets and then measure the similarity between two code snippets to predict whether they share common functionality.

\section{Experimental Design}
\label{sec:experimental_setup}

\subsection{Datasets}
\label{sec:dataset}

\subsubsection{Fault Datasets}

To ensure the thoroughness and validity of our research findings regarding fault localization, we have leveraged three widely used Java fault datasets: \textbf{Bears}~\cite{madeiral2019bears}, \textbf{Bugs.jar}~\cite{saha2018bugs}, and \textbf{Defects4J}~\cite{just2014defects4j}.
Since we focus on function-level, we perform two filtering steps on the original datasets to obtain functions, and the filtering results of each dataset are displayed in Table~\ref{tab:dataset_1}.

\begin{table}[htbp]
\centering
\caption{Statistic of the fault datasets}
\begin{center}
\resizebox{.8\linewidth}{!}
{
\begin{tabular}{lccccc}
\toprule
\textbf{Datasets} & \textbf{\# Buggy} & \textbf{\# Non-Buggy} & \textbf{\# Total} & \textbf{Buggy:Non-Buggy} \\
\midrule
Bears & 132 & 1,637 & 1,769 & 1:12.4 \\
Bugs.jar & 1,953 & 18,995 & 20,948 & 1:9.7 \\
Defects4J & 1,129 & 16,674 & 17,803 & 1:14.8 \\
\bottomrule
\end{tabular}
}
\label{tab:dataset_1}
\end{center}
\end{table}

\textbf{Step-1:} Each commit is considered as a version of a project.
We use the commit IDs to request commit histories, and for each commit, we extract the code changes before and after fixing a bug.
Finally, we use the code change information to obtain the buggy and fixed version of a function. 
Thus, we collect the following information for a project: buggy functions with their fixes and non-buggy functions.
In this step, we obtain the Bears dataset, consisting of 2,009 functions, the Bugs.jar dataset, consisting of 40,880 functions, and the Defects4J dataset, containing 31,423 functions.

\textbf{Step-2:} To clean and normalize the dataset, we start by removing duplicate functions. 
The three datasets are derived from various versions of projects (e.g., Defects4J extracted from 17 real-world Java projects), leading to a substantial number of duplicate functions extracted from different commits of the same project during step-1.
In this step, we finally obtain the Bears dataset, which comprises 1,769 functions, the Bugs.jar dataset, which comprises 20,948 functions, and the Defects4J dataset, which comprises 17,803 functions.

Regarding fault generation, Defects4J includes test suites for evaluating generated functions to determine if they are buggy functions.
In contrast, Bears and Bugs.jar do not include readily executable test suites.
Therefore, we instruct LLMs to automatically inject faults in the non-buggy functions of Defects4J.

\subsubsection{Clone Datasets}

We consider two datasets for the evaluation of clone detection: CodeXGLUE-POJ104~\cite{lu2021codexglue,mou2016convolutional} and CodeNet-Java250~\cite{puri2021codenet}. 
CodeXGLUE-POJ104 contains 104 programming challenges, and each has 500 C/C++ solutions from programmers. 
Code-XGLUE~\cite{lu2021codexglue} reconstruct it as a public benchmark by splitting the dataset into Training (64 challenges), Validation (16 challenges), and Testing (24 challenges) sets, making sure that there are no overlapped challenges between any two sets. 
CodeNet-Java250 contains 250 Java programming challenges from online judge websites, and each has 300 solutions from programmers. 
It splits the datasets into Training (125 challenges), Validation (62 challenges), and Testing (63 challenges) sets without overlapped challenges. 
The detailed statistics of these two datasets can be found in Table~\ref{tab:dataset_2}.

To generate additional clone data, we utilize the official prompts provided by HumanEval-X~\cite{zheng2023codegeex}.
HumanEval-X is a benchmark for evaluating the multilingual ability of code generative models. 
It consists of 820 high-quality human-crafted data samples (each with test cases) in Python, C++, Java, JavaScript, and Go, and can be used for various tasks, such as code generation and translation.

\begin{table}[htbp]
  \centering
  \caption{Statistic of the clone datasets}
  \resizebox{.8\linewidth}{!}{
    \begin{tabular}{lcccc}
    \toprule
    \textbf{Datasets} & {\textbf{Language}} & {\textbf{\# Training}} & {\textbf{\# Validation}} & {\textbf{\# Testing}} \\
    \midrule
    CodeXGLUE-POJ104 & C/C++ & 32,000 & 8,000 & 12,000 \\
    CodeNet-Java250 & Java & 37,500 & 18,600 & 18,900 \\
    \bottomrule
    \end{tabular}%
    }
  \label{tab:dataset_2}%
\end{table}%

\subsection{Evaluation Metrics}

To evaluate the effectiveness of fault localization, we consider the following metrics: Accuracy, Precision, Recall, F1-score, and FPR.

\textbf{Accuracy} evaluates the performance of how many code lines can be correctly labeled. 
It is calculated as:
$\frac{TP+ TN}{TP+FP+TN+FN}$.

\textbf{Precision} is the fraction of true buggy lines among the located ones. 
It is defined as:
$\frac{TP}{TP+FP}$.

\textbf{Recall} measures how many buggy lines can be correctly located. 
It is defined as: 
$\frac{TP}{TP+FN}$.

\textbf{F1-score} is a harmonic mean of $Precision$ and $Recall$ and can be calculated as:
$\frac{2 \times P \times R}{P + R}$.

\textbf{FPR} refers to the proportion of non-buggy ones that are predicted to be buggy.
It is defined as: $\frac{FP}{FP+TN}$.

As for clone detection, we adopt the MAP@R 
metric~\cite{lu2021codexglue,puri2021codenet}. 
MAP@R is a common metric to evaluate the quality of information retrieval, and it measures the average precision scores of a set of the top-R clone candidates presented in response to a query program.

\subsection{Implementation}
\label{sec:implementation}

We develop the fault generation and clone generation pipeline in Python, utilizing PyTorch~\cite{pytorch} implementations of LLMs (i.e., Magicoder 6.7B, DeepSeek-Coder 6.7B, OpenCodeInterpreter 6.7B, CodeLlama 7B, WizardCoder 7B, Phi-2 2.7B, Mistral 7B, and Llama 3 8B).
We use the Hugging Face API~\cite{huggingface} to load the model weights and generate outputs.
We also adhere to the best-practice guide~\cite{shieh2023best} for each prompt.
Our default setting for generation uses top p = 0.95, temperature = 1.
We use 10 samples for fault generation and 200 samples for clone generation.
Regarding pre-trained LMs, we utilize their publicly available source code and perform fine-tuning with the default parameters provided in their original code.
All these models are implemented using the PyTorch~\cite{pytorch} framework.
During fine-tuning, we employ the AdamW optimizer~\cite{loshchilov2017decoupled}, which is widely adopted to fine-tune Transformer-based models to optimize the parameters of LMs and LLMs. 
In our experiment, we set the maximum epochs of fault localization and clone detection to 20 and adopt the early stop mechanism to obtain better parameters.
The models (i.e., LMs and LLMs) with the best performance on the validation set are used for the evaluations.
Our evaluation is conducted on a 32-core workstation equipped with an Intel(R) Xeon(R) Platinum 8358P CPU @ 2.60GHz, 2TB RAM, and 8×NVIDIA A800 80GB GPU, running Ubuntu 20.04.6 LTS.

\section{Experimental Results}
\label{sec:results}

\begin{itemize}[leftmargin=*]

\item \textbf{RQ-1 Effectiveness of LLMs in Fault Generation and Clone Generation.
} {\em Can LLMs successfully inject faults into non-buggy functions and generate code clones? 
}

\item \textbf{RQ-2 Comparable Study of LLMs and LMs.} {\em How does the performance of LLMs in fault localization and clone detection compare against LMs?} 

\item \textbf{RQ-3 Enhancing LMs with LLM-generated Data.} {\em How well do LMs perform at two tasks when learning from LLM-generated faults and clones?}

\end{itemize}

\subsection{RQ-1: What are the Characteristics of the Generated Faults and Clones?}
\label{sec:rq1}

\noindent
\textbf{Objective}.
Recently, the exceptional performance of LLMs in code generation has garnered significant attention within the software engineering community. 
To overcome the challenge posed by the current scarcity and high cost of collecting real-world data, we leverage the powerful generation capabilities of LLMs to enhance pre-trained LMs. Specifically, we utilize LLMs to generate domain-specific data, thereby enhancing the performance of pre-trained models on specific tasks. 
Therefore, in this RQ, we aim to investigate and analyze the characteristics of faults and clones generated by LLMs in software code.

\noindent
\textbf{Experimental Setup}.
To answer this research question, we investigate the characteristics of the faults using the metrics listed below. 
We compute these metrics for all LLMs, as well as the average of these metrics for each project.

\begin{itemize}[leftmargin=*]

\item \textbf{Test fail}. 
This metric measures the number of confirmed faults (i.e., functions that fail to pass the test cases) generated by LLMs. 

\item \textbf{Test pass}. 
This metric measures the number of functions that pass the test cases.
Specifically, for fault generation, this metric indicates instances where the LLMs may fail to inject faults or the issue is not necessarily covered by the existing tests.
In contrast, for clone generation, it represents instances where the generated codes are correct.

\item \textbf{Time out}. 
This metric measures the number of functions that exceed the test timeout limit of 180 seconds.

\item \textbf{Other}. 
This metric measures the number of functions generated by LLMs that result in compilation failures or syntax errors.

\item \textbf{Lines involved in fault generation}. 
These metrics measure the average number of lines added, removed, or modified to generate the faults. 

\item \textbf{Edit distance}. 
This metric measures the minimum number of single-character edits (i.e., insertions, deletions, or substitutions) required to change one string into another.
Each code is represented as a string of characters to compute the Levenshtein edit distance~\cite{levenshtein1966binary}. 

\end{itemize}

For clone generation, we use the unbiased pass@k metric proposed in Codex~\cite{chen2021evaluating,zheng2023codegeex}. 
To evaluate pass@k, we generate $n \geq k$ samples per task and in this paper, we use $n = 200$ and $k \in \{1, 10, 100, 200\}$.

\begin{table}[htbp]
  \centering
  \caption{Statistics of the generated faults (RQ1)}
  {
    \begin{tabular}{lcccc}
    \toprule
    \textbf{Models} & \textbf{\# Test Fail} & \textbf{\# Test Pass} & \textbf{\# Time Out} & \textbf{\# Other} \\
    \midrule
    Magicoder & 22,809 & 56,850 & 1,565 & 23,747 \\
    DeepSeek-Coder & 23,136 & 51,072 & 1,468 & 35,698 \\
    OpenCodeInterpreter & 18,768 & 46,259 & 1,720 & 19,196 \\
    CodeLlama & 22,442 & 50,491 & 1,617 & 23,852 \\
    WizardCoder & 16,340 & 42,899 & 1,650 & 36,273 \\
    Phi-2 & 11,256 & 54,727 & 2,256 & 30,289 \\
    Mistral & 21,511 & 40,225 & 3,111 & 37,904 \\
    Llama 3 & 24,718 & 39,983 & 2,443 & 29,304 \\
    \midrule
    \textbf{All (Filtered)} & \textbf{112,709} & \textbf{-} & \textbf{-} & \textbf{-} \\
    \bottomrule
    \end{tabular}%
  }
  \label{tab:rq1_1}%
\end{table}%

\noindent
\textbf{Results.}
Table~\ref{tab:rq1_1} and Table~\ref{tab:rq1_2}
show the number of faults generated by LLMs and the characteristics of faults, while Table~\ref{tab:rq1_3} shows the characteristics of clones generated by LLMs.
Overall, \textbf{we find that LLMs possess the capability to generate faults and clones, enabling the creation of a vast array of non-duplicate data.}
We discuss the results from the characteristics of faults and clones, respectively.

\underline{\textbf{Characteristics of Faults.}}
As shown in Table~\ref{tab:rq1_1}, we can draw the following observations for fault generation: 
(1) All the LLMs can successfully generate faults for the non-buggy function in the subject projects.
Llama 3 generates 1,582$\sim$13,462 more faults compared to other LLMs.
\textbf{After filtering out duplicate functions, we ended up with 112,709 buggy functions.}
(2) During the fault generation process, LLMs often produce some invalid data. 
Specifically, the functions where LLMs fail to inject faults (i.e., \# Test Pass) and the functions that result in compilation failures or syntax errors (i.e., \# Other) account for a significant portion.

As shown in Table~\ref{tab:rq1_2}, we find that:
\textbf{(1) The generated faults mainly come from JacksonDatabind, Lang, and Math projects.} These three projects account for 46.6\% of all the generated faults.
(2) The average number of lines added, removed, or modified to generate the faults ranges from 0.48 to 2.07.
(3) The edit distance between different projects ranges from 42.67 to 75.05. 
Compared to other projects, JacksonXml's faults have higher edit distance values. 
This indicates that the changes made by LLMs in the JacksonXml project differ more significantly from the original code.

\begin{table}[htbp]
  \centering
  \caption{Characteristics of the generated faults (RQ1)}
  \resizebox{.8\linewidth}{!}
  {
    \begin{tabular}{lccccc}
    \toprule
    \textbf{Projects} & \textbf{\# Test Fail} & \textbf{Add} & \textbf{Remove} & \textbf{Modify} & \textbf{Edit Distance} \\
    \midrule
    Chart & 6,815 & 0.82 & 1.85 & 0.90 & 66.55 \\
    Cli & 3,900 & 1.05 & 1.78 & 1.08 & 61.87 \\
    Closure & 5,061 & 1.57 & 1.32 & 1.19 & 62.10 \\
    Codec & 2,824 & 0.83 & 0.87 & 1.06 & 47.24 \\
    Collections & 1,314 & 0.88 & 1.51 & 0.95 & 56.30 \\
    Compress & 7,426 & 0.89 & 1.35 & 1.01 & 60.96 \\
    Csv & 2,019 & 0.80 & 0.78 & 0.98 & 50.10 \\
    Gson & 577 & 1.62 & 0.82 & 1.04 & 52.47 \\
    JacksonCore & 7,446 & 0.85 & 1.57 & 1.07 & 58.90 \\
    JacksonDatabind & 12,927 & 0.68 & 1.28 & 1.02 & 60.56 \\
    JackonXml & 778 & 0.83 & 2.07 & 0.81 & 75.05 \\
    Jsoup & 9,845 & 0.77 & 0.71 & 1.03 & 43.48 \\
    JxPath & 2,859 & 1.06 & 1.70 & 1.35 & 68.23 \\
    Lang & 19,726 & 1.02 & 0.99 & 1.11 & 51.47 \\
    Math & 19,912 & 0.66 & 0.84 & 1.11 & 42.67 \\
    Mockito & 2,978 & 0.48 & 1.44 & 0.95 & 74.56 \\
    Time & 6,302 & 0.58 & 0.91 & 0.98 & 49.84 \\
    \bottomrule
    \end{tabular}%
  }
  \label{tab:rq1_2}%
\end{table}%

\underline{\textbf{Characteristics of Clones.}}
Table~\ref{tab:rq1_3} presents the characteristics of the generated clones. 
In general, \textbf{we find that the code generation capabilities of LLMs are stronger for Java than for C++.}
For example, LLMs achieve a higher average pass@200 score in HumanEval-X-Java compared to HumanEval-X-C++ (i.e., 55.79\% v.s. 41.08\%).
Furthermore, after filtering out duplicate data, LLMs generated 5,000 clones in HumanEval-X-C++ and 10,546 clones in HumanEval-X-Java.

\begin{table*}[htbp]
  \centering
  \caption{Characteristics of the generated clones (RQ1)}
  \resizebox{\linewidth}{!}
  {
    \begin{tabular}{l|ccccc|ccccc}
    \toprule
    \multirow{1.5}[4]{*}{\textbf{Models}} & \multicolumn{5}{c|}{\textbf{HumanEval-X-C++}} & \multicolumn{5}{c}{\textbf{HumanEval-X-Java}} \\
    \cmidrule{2-11} & \textbf{pass@1} & \textbf{pass@10} & \textbf{pass@100} & \textbf{pass@200} & \textbf{\# Test Pass} & \textbf{pass@1} & \textbf{pass@10} & \textbf{pass@100} & \textbf{pass@200} & \textbf{\# Test Pass} \\
    \midrule
    Magicoder & 11.62  & 44.77  & 67.57  & 70.12  & 3,812 & 18.50  & 50.50  & 66.43  & 68.90  & 6,067 \\
    DeepSeek-Coder & 1.36  & 10.98  & 45.01  & 57.93  & 446 & 2.77  & 20.98  & 55.12  & 60.37  & 910 \\
    OpenCodeInterpreter & 1.23  & 10.28  & 42.01  & 51.83  & 403 & 4.99  & 30.76  & 60.45  & 64.63  & 1,638 \\
    CodeLlama & 1.68  & 11.98  & 36.13  & 45.73  & 552 & 9.73  & 33.14  & 53.64  & 59.15  & 3,191 \\
    WizardCoder & 0.36  & 3.32  & 19.15  & 26.83  & 119 & 3.01  & 19.30  & 45.49  & 50.00  & 988 \\
    Phi-2 & 0.25  & 2.35  & 14.55  & 20.73  & 82 & 0.57  & 5.13  & 24.13  & 31.10  & 188 \\
    Mistral & 0.83  & 6.18  & 27.49  & 34.76  & 272 & 5.13  & 22.60  & 41.02  & 46.34  & 1,684 \\
    Llama 3 & 0.22  & 2.07  & 13.71  & 21.34  & 72 & 3.78  & 25.05  & 59.64  & 65.85  & 1,239 \\
    \midrule
    \textbf{All (Filtered)} & - & - & - & - & \textbf{5,000} & - & - & - & - & \textbf{10,546} \\
    \bottomrule
  \end{tabular}%
  }
  \label{tab:rq1_3}%
\end{table*}

\intuition{
\textbf{Finding 1}: 
LLMs are capable of generating a substantial amount of faults and clones, significantly enhancing the data available for fault localization and clone detection.
}

\subsection{RQ-2: How does Directly
Applying LLMs for Two Tasks Compare Against LMs?}
\label{sec:rq2}

\noindent
\textbf{Objective}.
Benefiting from the powerful representation capability of deep neural networks, many pre-trained LMs have been proposed~\cite{feng2020codebert,guo2022unixcoder,guo2020graphcodebert,ahmad2021unified,wang2021codet5,wang2023codet5+} and they treat the source code in different ways.
Meanwhile, recently, LLMs~\cite{deepseek-coder,roziere2023code,luo2023wizardcoder,wei2023magicoder,zheng2024opencodeinterpreter,meta2024llama,jiang2023mistral,phi} have attracted much attention since their powerful ability can be easily adapted to various types of downstream tasks~\cite{bang2023multitask,ouyang2022training}.
However, the efficiency and overhead of LLMs and pre-trained LMs in fault localization and clone detection have not been systematically compared.
Considering these issues, we aim to conduct an extensive study to comprehensively compare LLMs and pre-trained LMs.

\begin{table*}[!htbp]
  \centering
  \caption{Results for LMs and LLMs in fault localization (RQ2)}
  \resizebox{\linewidth}{!}
  {
    \begin{tabular}{l|ccccc|ccccc}
    \toprule
    \multirow{1.5}[4]{*}{\textbf{Models}} & \multicolumn{5}{c|}{\textbf{Defects4J}} & \multicolumn{5}{c}{\textbf{Real-World Datasets}} \\
\cmidrule{2-11}      & \textbf{F1-score} & \textbf{Recall} & \textbf{Precision} & \textbf{Accuracy} & \textbf{FPR} & \textbf{F1-score} & \textbf{Recall} & \textbf{Precision} & \textbf{Accuracy} & \textbf{FPR} \\
    \midrule
    BERT & 0.341  & 0.402  & 0.297  & 0.843  & 0.108  & 0.201  & 0.150  & 0.303  & 0.841  & 0.053  \\
    RoBERTa & 0.394  & 0.412  & 0.378  & 0.876  & 0.073  & 0.286  & \textbf{0.258} & 0.322  & 0.836  & 0.079  \\
    CodeBERT & \textbf{0.428} & \textbf{0.434} & 0.421  & 0.887  & 0.065  & 0.269  & 0.217  & 0.353  & 0.849  & 0.058  \\
    GraphCodeBERT & 0.384  & 0.353  & 0.421  & 0.889  & 0.053  & \textbf{0.290} & 0.241  & 0.364  & 0.849  & 0.062  \\
    UniXcoder & 0.409  & 0.396  & 0.423  & 0.889  & 0.058  & 0.220  & 0.162  & 0.343  & 0.856  & 0.044  \\
    PLBART & 0.395  & 0.348  & \textbf{0.456} & \textbf{0.894} & \textbf{0.046}  & 0.244  & 0.169  & \textbf{0.443} & \textbf{0.865} & \textbf{0.032} \\
    CodeT5 & 0.392  & 0.367  & 0.422  & 0.889  & 0.055  & 0.250  & 0.191  & 0.363  & 0.856  & 0.048  \\
    CodeT5+ & 0.392  & 0.360  & 0.431  & 0.891  & 0.052 & 0.241  & 0.177  & 0.395  & 0.863  & 0.038  \\
    \midrule
    \rowcolor{lightgray}\textbf{Fine-Tuning Setting} &&&&&&&&&& \\
    Magicoder & 0.359  & 0.331  & 0.391  & 0.885  & 0.056  & 0.236  & 0.182  & 0.337  & 0.850  & 0.052  \\
    CodeLlama & 0.429  & 0.331  & 0.608  & 0.914  & 0.023  & 0.208  & 0.132  & \textbf{0.491} & 0.872  & 0.020  \\
    WizardCoder & 0.404  & 0.331  & 0.517  & 0.904  & 0.034  & 0.217  & 0.148  & 0.409  & 0.864  & 0.031  \\
    DeepSeek-Coder & 0.376  & 0.324  & 0.449  & 0.895  & 0.043  & 0.233  & 0.168  & 0.383  & 0.860  & 0.039  \\
    OpenCodeInterpreter & 0.393  & 0.309  & 0.539  & 0.907  & 0.029  & 0.202  & 0.134  & 0.414  & 0.866  & 0.028  \\
    Phi-2 & 0.358  & \textbf{0.456} & 0.295  & 0.841  & 0.117  & \textbf{0.264} & \textbf{0.299} & 0.237  & 0.788  & 0.141  \\
    Mistral & 0.310  & 0.309  & 0.311  & 0.866  & 0.074  & 0.162  & 0.126  & 0.225  & 0.834  & 0.063  \\
    Llama 3 & \textbf{0.453} & 0.355  & \textbf{0.625} & \textbf{0.919} & \textbf{0.022} & 0.164  & 0.098  & 0.489  & \textbf{0.878} & \textbf{0.014} \\
    \midrule
    \rowcolor{lightgray}\textbf{Few-Shot Setting} &&&&&&&&&& \\
    Magicoder & 0.151  & 0.425  & 0.092  & 0.617  & 0.366  & 0.197  & 0.425  & 0.128  & 0.603  & 0.374  \\
    CodeLlama & \textbf{0.177} & \textbf{0.603} & 0.104  & 0.547  & 0.458  & \textbf{0.220} & \textbf{0.558} & 0.137  & 0.547  & 0.455  \\
    WizardCoder & 0.170  & 0.582  & 0.100  & 0.541  & 0.463  & 0.218  & 0.528  & 0.137  & 0.567  & 0.429  \\
    DeepSeek-Coder & 0.171  & 0.452  & 0.105  & 0.647  & 0.336  & 0.195  & 0.389  & 0.130  & 0.633  & 0.335  \\
    OpenCodeInterpreter & 0.169  & 0.486  & 0.103  & 0.616  & 0.373  & 0.214  & 0.483  & 0.137  & 0.593  & 0.392  \\
    Phi-2 & 0.112  & 0.096  & \textbf{0.133} & \textbf{0.877} & \textbf{0.055} & 0.108  & 0.083  & 0.153  & \textbf{0.843} & \textbf{0.059} \\
    Mistral & 0.098  & 0.096  & 0.100  & 0.857  & 0.076  & 0.140  & 0.111  & \textbf{0.188} & \textbf{0.843} & 0.062  \\
    Llama 3 & 0.158  & 0.280  & 0.110  & 0.762  & 0.196  & 0.173  & 0.251  & 0.132  & 0.728  & 0.211  \\
    \bottomrule
    \end{tabular}%
  }
  \label{tab:rq2_fault_localization}%
\end{table*}%

\noindent
\textbf{Experimental Setup}.
We consider eight LLMs (i.e., Magicoder~\cite{wei2023magicoder}, OpenCodeInterpreter~\cite{zheng2024opencodeinterpreter}, DeepSeek-Coder~\cite{deepseek-coder}, CodeLlama~\cite{roziere2023code}, WizardCoder~\cite{luo2023wizardcoder}, Llama 3~\cite{meta2024llama}, Mistral~\cite{jiang2023mistral}, and Phi-2~\cite{phi}) and eight LMs (i.e., BERT~\cite{devlin2018bert}, RoBERTa~\cite{liu2019roberta}, CodeBERT~\cite{feng2020codebert}, GraphCodeBERT~\cite{guo2020graphcodebert}, UniXcoder~\cite{guo2022unixcoder}, PLBART~\cite{ahmad2021unified}, CodeT5~\cite{wang2021codet5}, and CodeT5+~\cite{wang2023codet5+}).

Since we evaluate two distinct downstream tasks, the datasets and data processing methods used varied accordingly. 
For the fault localization task, we utilize all buggy functions from the Defects4J dataset.
We divide these functions into training, validation, and testing sets in an 8:1:1 ratio.
Additionally, we extend our evaluation by training and validating on the Detects4J dataset, but testing on the Bears and Bugs.jar datasets.
This experiment aims to showcase the localization capabilities of different models in locating unknown real-world faults.
Regarding the clone detection task, we use datasets divided in previous works~\cite{guo2022unixcoder,ding2023concord}, including CodeXGLUE-POJ104 and CodeNet-Java250.

For LMs, we use the training set to fine-tune the models.
In contrast, for LLMs, we use both fine-tuning and few-shot settings.
The purpose of the few-shot setting is to evaluate the performance of LLMs under limited computing resources and time constraints.

In terms of fine-tuning setting, we adopt the parameters and configurations described in Section~\ref{sec:implementation} to fine-tune both LLMs and LMs, and then evaluate them on the testing set. 
For the few-shot setting, we use the testing set for the evaluation and instruct LLMs with the following task descriptions to tell it to act as a fault locator.

\intuition{
\textbf{Fault Localization:} 
I will provide you a buggy Java code snippet and please locate buggy lines.
}

The clone detection metric MAP@R requires assessing the similarity between a code snippet and all other code snippets in the dataset. 
However, due to the token limitations of LLMs, we are unable to implement clone detection in a few-shot setting.

In order to comprehensively compare the performance among LLMs and LMs, we consider six widely used performance measures (i.e., Precision, Recall, F1-score, Accuracy, FPR, and MAP@R).

\begin{table}[htbp]
  \centering
  \caption{Results for LMs and LLMs in clone detection (RQ2)}
  \resizebox{.8\linewidth}{!}
  {
    \begin{tabular}{lcc}
    \toprule
    \textbf{Models} & \textbf{CodeXGLUE-POJ104} & \textbf{CodeNet-Java250} \\
    \midrule
    BERT & 78.65\% & 73.62\% \\
    RoBERTa & 81.65\% & 77.13\% \\
    CodeBERT & 84.93\% & 80.98\% \\
    GraphCodeBERT & 83.55\% & 83.09\% \\
    UniXcoder & 90.50\% & 83.45\% \\
    PLBART & 85.34\% & 82.79\% \\
    CodeT5 & 90.46\% & \textbf{85.58\%} \\
    CodeT5+ & \textbf{90.52\%} & 82.35\% \\
    \midrule
    \rowcolor{lightgray}\textbf{Fine-Tuning Setting} & & \\
    Magicoder & 54.62\% & 48.06\% \\
    CodeLlama & 56.52\% & 48.42\% \\
    WizardCoder & 56.49\% & 48.38\% \\
    DeepSeek-Coder & 54.68\% & 48.06\% \\
    OpenCodeInterpreter & 56.49\% & 48.04\% \\
    Phi-2 & \textbf{59.97\%} & \textbf{51.13\%} \\
    Mistral & 54.98\% & 48.12\% \\
    Llama 3 & 58.38\% & 50.65\% \\
    \bottomrule
    \end{tabular}%
  }
  \label{tab:rq2_clone_detection}%
\end{table}%

\noindent
\textbf{Results.}
Table~\ref{tab:rq2_fault_localization} and Table~\ref{tab:rq2_clone_detection}
show the results of fault localization and clone detection, while Table~\ref{tab:rq2_efficiency} shows the average time cost and GPU memory cost for fine-tuning different types of models for one epoch.
Overall, \textbf{we find that fine-tuned LLMs exhibit slightly superior performance in fault localization compared to LMs. However, this comes at a significant cost in terms of time and computational resources.}
We discuss the results from the aspects of effectiveness and efficiency, respectively.

\textbf{\underline{Comparison of Effectiveness.}}
As shown in Table~\ref{tab:rq2_fault_localization}, we can draw the following observations for fault localization: 
(1) There is no significant difference between encoder models and encoder-decoder models.
After fine-tuning, all LMs show promising performance.
Among them, PLBART exhibits the best performance in terms of Precision, Accuracy, and FPR on both the Defects4J and real-world datasets.
\textbf{(2) Under the fine-tuning setting, although the best performance of LLMs surpasses that of LMs, the difference was not significant.}
For example, on Defects4J, the F1-score increases from 0.428 to 0.453, the Recall increases from 0.434 to 0.456, the Precision increases from 0.456 to 0.625, the Accuracy increases from 0.894 to 0.919, and the FPR decreases from 0.046 to 0.022.
(3) Under the few-shot setting, LLMs show very poor fault localization capability, with much lower precision and F1-score compared to LLMs and LMs under fine-tuning settings.
This indicates the necessity of fine-tuning for specific datasets in specific task scenarios, as LLMs cannot be universally applied to any scenario.
(4) Regardless of whether they are LMs or LLMs, their performance on the real-world datasets after being fine-tuned on the Defects4J dataset is very poor, even comparable to the Few-Shot Setting.
This highlights the importance of domain-specific data and indicates that the current limited datasets are insufficient for fine-tuning models effectively.

As shown in Table~\ref{tab:rq2_clone_detection}, \textbf{we find that LMs outperform LLMs significantly in clone detection.}
For instance, CodeT5+ achieves a MAP@R of 90.52\% on POJ104 dataset and CodeT5 achieves a MAP@R of 85.58\% on Java250 dataset.
In contrast, the highest-performing LLM, Phi-2, only achieves a MAP@R of 59.97\% and 51.13\% on POJ104 and Java250 respectively.

\textbf{\underline{Comparison of Efficiency.}}
Table~\ref{tab:rq2_efficiency} presents the details of time cost and GPU memory cost for different models in our study. 
\textbf{We find that fine-tuning LLMs requires significantly more time and computational resources.}
On the whole, fine-tuning encoder-only LMs and encoder-decoder LMs requires only about 35.2\% to 81.3\% of the time cost needed to fine-tune LLMs.
Additionally, the GPU memory cost of LMs is much lower than that of LLMs.
This is mainly because LMs typically have 110M to 220M parameters, while LLMs have 2.7B to 8B parameters.

\begin{table}[htbp]
  \centering
  \caption{Average time cost and GPU memory cost for fine-tuning different models for one epoch (RQ2)}
  \resizebox{.7\linewidth}{!}
  {
    \begin{tabular}{l|cc|cc}
    \toprule
    \multirow{1.5}[4]{*}{\textbf{Models}} & \multicolumn{2}{c|}{\textbf{Fault Localization}} & \multicolumn{2}{c}{\textbf{Clone Detection}} \\
    \cmidrule{2-5} & \textbf{Time} & \textbf{GPU Memory} & \textbf{Time} & \textbf{GPU Memory} \\
    \midrule
    LMs & 1m 14s & 3,189M  & 46m 53s  & 11,551M  \\
    LLMs & 3m 30s & 64,031M  & 57m 40s & 53,253M  \\
    \bottomrule
    \end{tabular}%
  }
  \label{tab:rq2_efficiency}%
\end{table}%

\intuition{
\textbf{Finding 2}: Investing significant time and computational resources in fine-tuning LLMs, only to achieve minor performance improvements compared to LMs (and sometimes even inferior performance, as observed in clone detection), is not considered worthwhile.
}

\subsection{RQ-3: Does LLM-generated Data Improve the Performance of LMs?}
\label{sec:rq3}

\noindent
\textbf{Objective.}
In RQ2, we find that fine-tuning LLMs requires more time and computational resources but does not necessarily yield outstanding performance improvements. 
Furthermore, current LMs are constrained by the lack of domain-specific data, and collecting domain data is costly. 
By pre-training on large amounts of open-source code snippets, LLMs have the ability to generate code directly based on the surrounding context. 
Therefore, in this RQ, we aim to investigate whether LLMs can transfer their learned knowledge to LMs. 
Specifically, we explore if LLMs can generate data (e.g., fault data and clone data) for specific tasks to enhance the performance of LMs on downstream tasks.

\noindent
\textbf{Experimental Setup.}
To evaluate the effectiveness of using the LLM-generated data to improve the LMs. 
We adopt the experiment settings described in Section~\ref{sec:implementation} and perform the evaluation on eight LMs (i.e., BERT~\cite{devlin2018bert}, RoBERTa~\cite{liu2019roberta}, CodeBERT~\cite{feng2020codebert}, GraphCodeBERT~\cite{guo2020graphcodebert}, UniXcoder~\cite{guo2022unixcoder}, PLBART~\cite{ahmad2021unified}, CodeT5~\cite{wang2021codet5}, and CodeT5+~\cite{wang2023codet5+}). 
In terms of fault localization, we use the score to rank the generated data (refer to Section~\ref{sec:selection} for more detail) and select 10\%, 20\%, 50\%, and 100\% to add to the training set of Defects4J.
We also extend our evaluation by testing on the Bears and Bugs.jar datasets, which are considered to be unknown real-world datasets (no intersection with the LLM-generated faults).
For clone detection, we first calculate the average edit distance for all clone instances under each generation task. 
Then, we compute the edit distance between each clone instance and other clone instances. 
We select all clone instances with an edit distance greater than the average edit distance and add them to the training set of CodeXGLUE-POJ104 and CodeNet-Java250.
Additionally, in order to comprehensively evaluate whether the new training set enhances the performance of the LMs, we use the same metrics as RQ-2.

\begin{figure}[htbp]
    \centering
    \includegraphics[width=.6\linewidth]{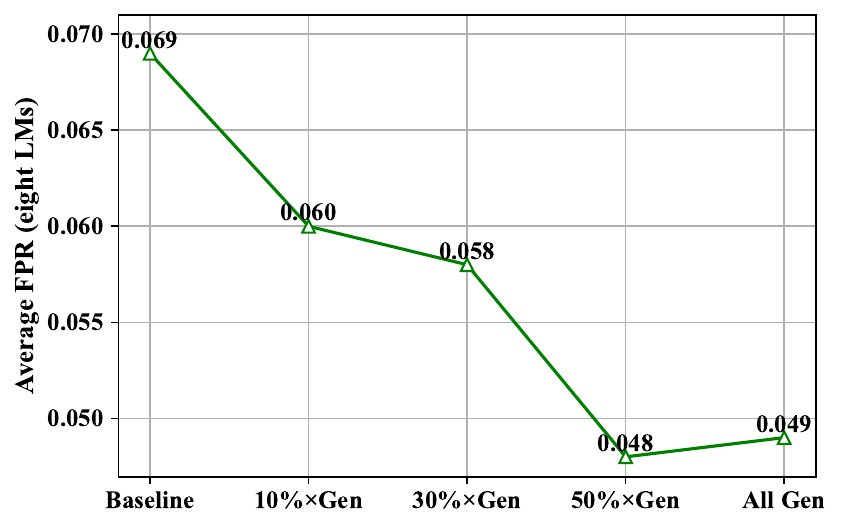}
    \caption{Average decrease (eight LMs) of FPR in fault localization (RQ3)}
    \label{fig:rq3}
\end{figure}

\noindent
\textbf{Results.}
Table~\ref{tab:rq3_fault_localization}, Table~\ref{tab:rq3_additional}, and Fig.~\ref{fig:rq3} illustrate the performance improvements of LMs in fault localization after the incorporation of data generated by LLMs, while Table~\ref{tab:rq3_clone_detection} shows the enhancements of these LMs in the area of code clone detection.
Overall, \textbf{we find that LLM-generated data can significantly enhance the performance of LMs in both fault localization and code clone detection tasks.}
We discuss the results from the aspects of fault localization and clone detection, respectively.

\begin{table*}[htbp]
  \centering
  \caption{Increase of F1-score, Recall, Precision, and Accuracy in fault localization (RQ3)}
  \resizebox{\linewidth}{!}
  {
    \begin{tabular}{l|ccccc|ccccc}
    \toprule
    \multirow{1.5}[4]{*}{\textbf{Models}} & \multicolumn{5}{c|}{\textbf{F1-score}} & \multicolumn{5}{c}{\textbf{Recall}} \\
    \cmidrule{2-11} & \textbf{Baseline} & \textbf{10\%×Generated} & \textbf{30\%×Generated} & \textbf{50\%×Generated} & \textbf{All Generated} & \textbf{Baseline} & \textbf{10\%×Generated} & \textbf{30\%×Generated} & \textbf{50\%×Generated} & \textbf{All Generated} \\
    \midrule
    BERT & 0.341 & \cellcolor[HTML]{E3F2D9}\textbf{0.354 (↑3.81\%)} & \cellcolor[HTML]{C9E4B4}\textbf{0.400 (↑17.30\%)} & \cellcolor[HTML]{E3F2D9}\textbf{0.344 (↑0.88\%)} & \cellcolor[HTML]{E3F2D9}\textbf{0.357 (↑4.69\%)} & 0.402 & 0.331 (↓17.66\%) & 0.394 (↓1.99\%) & 0.307 (↓23.63\%) & 0.315 (↓21.64\%) \\
    RoBERTa & 0.394 & \cellcolor[HTML]{E3F2D9}\textbf{0.406 (↑3.05\%)} & \cellcolor[HTML]{C9E4B4}\textbf{0.434 (↑10.15\%)} & \cellcolor[HTML]{E3F2D9}\textbf{0.432 (↑9.64\%)} & 0.382 (↓3.05\%) & 0.412 & 0.404 (↓1.94\%) & 0.397 (↓3.64\%) & \cellcolor[HTML]{E3F2D9}\textbf{0.434 (↑5.34\%)} & 0.309 (↓25.00\%) \\
    CodeBERT & 0.428 & \cellcolor[HTML]{E3F2D9}\textbf{0.443 (↑3.50\%)} & \cellcolor[HTML]{C9E4B4}\textbf{0.478 (↑11.68\%)} & \cellcolor[HTML]{E3F2D9}\textbf{0.461 (↑7.71\%)} & 0.428 ($\sim$0.00\%) & 0.434 & \cellcolor[HTML]{ADD88D}\textbf{0.559 (↑28.80\%)} & 0.397 (↓8.53\%) & \cellcolor[HTML]{C9E4B4}\textbf{0.500 (↑15.21\%)} & 0.360 (↓17.05\%) \\
    GraphCodeBERT & 0.384 & \cellcolor[HTML]{C9E4B4}\textbf{0.429 (↑11.72\%)} & \cellcolor[HTML]{588E31}\textbf{0.550 (↑43.23\%)} & \cellcolor[HTML]{588E31}\textbf{0.524 (↑36.46\%)} & \cellcolor[HTML]{C9E4B4}\textbf{0.454 (↑18.23\%)} & 0.353 & \cellcolor[HTML]{C9E4B4}\textbf{0.397 (↑12.46\%)} & \cellcolor[HTML]{588E31}\textbf{0.522 (↑47.88\%)} & \cellcolor[HTML]{588E31}\textbf{0.559 (↑58.36\%)} & \cellcolor[HTML]{E3F2D9}\textbf{0.382 (↑8.22\%)} \\
    UniXcoder & 0.409 & \cellcolor[HTML]{C9E4B4}\textbf{0.486 (↑18.83\%)} & \cellcolor[HTML]{C9E4B4}\textbf{0.457 (↑11.74\%)} & \cellcolor[HTML]{ADD88D}\textbf{0.517 (↑26.41\%)} & \cellcolor[HTML]{C9E4B4}\textbf{0.460 (↑12.47\%)} & 0.396 & \cellcolor[HTML]{ADD88D}\textbf{0.511 (↑29.04\%)} & \cellcolor[HTML]{E3F2D9}\textbf{0.417 (↑5.30\%)} & \cellcolor[HTML]{ADD88D}\textbf{0.504 (↑27.27\%)} & \cellcolor[HTML]{C9E4B4}\textbf{0.453 (↑14.39\%)} \\
    PLBART & 0.395 & \cellcolor[HTML]{C9E4B4}\textbf{0.436 (↑10.38\%)} & \cellcolor[HTML]{C9E4B4}\textbf{0.460 (↑16.46\%)} & \cellcolor[HTML]{C9E4B4}\textbf{0.457 (↑15.70\%)} & \cellcolor[HTML]{C9E4B4}\textbf{0.441 (↑11.65\%)} & 0.348 & \cellcolor[HTML]{C9E4B4}\textbf{0.393 (↑12.93\%)} & \cellcolor[HTML]{C9E4B4}\textbf{0.400 (↑14.94\%)} & \cellcolor[HTML]{C9E4B4}\textbf{0.393 (↑12.93\%)} & \cellcolor[HTML]{C9E4B4}\textbf{0.415 (↑19.25\%)} \\
    CodeT5 & 0.392 & \cellcolor[HTML]{ADD88D}\textbf{0.475 (↑21.17\%)} & \cellcolor[HTML]{588E31}\textbf{0.536 (↑36.73\%)} & \cellcolor[HTML]{588E31}\textbf{0.537 (↑36.99\%)} & \cellcolor[HTML]{C9E4B4}\textbf{0.457 (↑16.58\%)} & 0.367 & \cellcolor[HTML]{E3F2D9}\textbf{0.374 (↑1.91\%)} & \cellcolor[HTML]{588E31}\textbf{0.540 (↑47.14\%)} & \cellcolor[HTML]{588E31}\textbf{0.496 (↑35.15\%)} & \cellcolor[HTML]{ADD88D}\textbf{0.475 (↑29.43\%)} \\
    CodeT5+ & 0.392 & \cellcolor[HTML]{C9E4B4}\textbf{0.447 (↑14.03\%)} & 0.387 (↓1.28\%) & 0.346 (↓11.73\%) & \cellcolor[HTML]{E3F2D9}\textbf{0.396 (↑1.02\%)} & 0.360 & \cellcolor[HTML]{588E31}\textbf{0.475 (↑31.94\%)} & \cellcolor[HTML]{588E31}\textbf{0.568 (↑57.78\%)} & 0.237 (↓34.17\%) & \cellcolor[HTML]{C9E4B4}\textbf{0.396 (↑10.00\%)} \\
    \midrule
    \multirow{1.5}[4]{*}{\textbf{Models}} & \multicolumn{5}{c|}{\textbf{Precision}} & \multicolumn{5}{c}{\textbf{Accuracy}} \\
    \cmidrule{2-11} & \textbf{Baseline} & \textbf{10\%×Generated} & \textbf{30\%×Generated} & \textbf{50\%×Generated} & \textbf{All Generated} & \textbf{Baseline} & \textbf{10\%×Generated} & \textbf{30\%×Generated} & \textbf{50\%×Generated} & \textbf{All Generated} \\
    \midrule
    BERT & 0.297 & \cellcolor[HTML]{ADD88D}\textbf{0.382 (↑28.62\%)} & \cellcolor[HTML]{588E31}\textbf{0.407 (↑37.04\%)} & \cellcolor[HTML]{588E31}\textbf{0.390 (↑31.31\%)} & \cellcolor[HTML]{588E31}\textbf{0.412 (↑38.72\%)} & 0.843 & \cellcolor[HTML]{E3F2D9}\textbf{0.878 (↑4.15\%)} & \cellcolor[HTML]{E3F2D9}\textbf{0.880 (↑4.39\%)} & \cellcolor[HTML]{E3F2D9}\textbf{0.881 (↑4.51\%)} & \cellcolor[HTML]{E3F2D9}\textbf{0.885 (↑4.98\%)} \\
    RoBERTa & 0.378 & \cellcolor[HTML]{E3F2D9}\textbf{0.407 (↑7.67\%)} & \cellcolor[HTML]{ADD88D}\textbf{0.478 (↑26.46\%)} & \cellcolor[HTML]{C9E4B4}\textbf{0.431 (↑14.02\%)} & \cellcolor[HTML]{588E31}\textbf{0.500 (↑32.28\%)} & 0.876 & \cellcolor[HTML]{E3F2D9}\textbf{0.884 (↑0.91\%)} & \cellcolor[HTML]{E3F2D9}\textbf{0.899 (↑2.63\%)} & \cellcolor[HTML]{E3F2D9}\textbf{0.889 (↑1.48\%)} & \cellcolor[HTML]{E3F2D9}\textbf{0.902 (↑2.97\%)} \\
    CodeBERT & 0.421 & 0.367 (↓12.83\%) & \cellcolor[HTML]{588E31}\textbf{0.600 (↑42.52\%)} & \cellcolor[HTML]{E3F2D9}\textbf{0.428 (↑1.66\%)} & \cellcolor[HTML]{ADD88D}\textbf{0.527 (↑25.18\%)} & 0.887 & 0.863 (↓2.71\%) & \cellcolor[HTML]{E3F2D9}\textbf{0.915 (↑3.16\%)} & 0.886 (↓0.11\%) & \cellcolor[HTML]{E3F2D9}\textbf{0.906 (↑2.14\%)} \\
    GraphCodeBERT & 0.421 & \cellcolor[HTML]{C9E4B4}\textbf{0.466 (↑10.69\%)} & \cellcolor[HTML]{588E31}\textbf{0.582 (↑38.24\%)} & \cellcolor[HTML]{C9E4B4}\textbf{0.494 (↑17.34\%)} & \cellcolor[HTML]{588E31}\textbf{0.559 (↑32.78\%)} & 0.889 & \cellcolor[HTML]{E3F2D9}\textbf{0.897 (↑0.90\%)} & \cellcolor[HTML]{E3F2D9}\textbf{0.917 (↑3.15\%)} & \cellcolor[HTML]{E3F2D9}\textbf{0.901 (↑1.35\%)} & \cellcolor[HTML]{E3F2D9}\textbf{0.910 (↑2.36\%)} \\
    UniXcoder & 0.423 & \cellcolor[HTML]{E3F2D9}\textbf{0.464 (↑9.69\%)} & \cellcolor[HTML]{C9E4B4}\textbf{0.504 (↑19.15\%)} & \cellcolor[HTML]{ADD88D}\textbf{0.530 (↑25.30\%)} & \cellcolor[HTML]{C9E4B4}\textbf{0.467 (↑10.40\%)} & 0.889 & \cellcolor[HTML]{E3F2D9}\textbf{0.895 (↑0.67\%)} & \cellcolor[HTML]{E3F2D9}\textbf{0.903 (↑1.57\%)} & \cellcolor[HTML]{E3F2D9}\textbf{0.908 (↑2.14\%)} & \cellcolor[HTML]{E3F2D9}\textbf{0.897 (↑0.90\%)} \\
    PLBART & 0.456 & \cellcolor[HTML]{E3F2D9}\textbf{0.491 (↑7.68\%)} & \cellcolor[HTML]{C9E4B4}\textbf{0.540 (↑18.42\%)} & \cellcolor[HTML]{C9E4B4}\textbf{0.546 (↑19.74\%)} & \cellcolor[HTML]{E3F2D9}\textbf{0.471 (↑3.29\%)} & 0.894 & \cellcolor[HTML]{E3F2D9}\textbf{0.899 (↑0.56\%)} & \cellcolor[HTML]{E3F2D9}\textbf{0.907 (↑1.45\%)} & \cellcolor[HTML]{E3F2D9}\textbf{0.907 (↑1.45\%)} & \cellcolor[HTML]{E3F2D9}\textbf{0.896 (↑0.22\%)} \\
    CodeT5 & 0.422 & \cellcolor[HTML]{588E31}\textbf{0.650 (↑54.03\%)} & \cellcolor[HTML]{588E31}\textbf{0.532 (↑26.07\%)} & \cellcolor[HTML]{588E31}\textbf{0.585 (↑38.63\%)} & \cellcolor[HTML]{E3F2D9}\textbf{0.440 (↑4.27\%)} & 0.889 & \cellcolor[HTML]{E3F2D9}\textbf{0.919 (↑3.37\%)} & \cellcolor[HTML]{E3F2D9}\textbf{0.908 (↑2.14\%)} & \cellcolor[HTML]{E3F2D9}\textbf{0.916 (↑3.04\%)} & 0.889 ($\sim$0.00\%) \\
    CodeT5+ & 0.431 & 0.423 (↓1.86\%) & 0.294 (↓31.79\%) & \cellcolor[HTML]{588E31}\textbf{0.635 (↑47.33\%)} & 0.396 (↓8.12\%) & 0.891 & 0.885 (↓0.67\%) & 0.824 (↓7.52\%) & \cellcolor[HTML]{E3F2D9}\textbf{0.912 (↑2.36\%)} & 0.882 (↓1.01\%) \\
    \bottomrule
    \end{tabular}%
  }
  \label{tab:rq3_fault_localization}%
\end{table*}%

\textbf{\underline{Improvement of Fault Localization.}}
As shown in Table~\ref{tab:rq3_fault_localization} and Fig.~\ref{fig:rq3}, we can
draw the following observations:

\textbf{LLM-generated data brings noticeable improvements to LMs.}
Overall, the addition of generated data reduces the model's average FPR.
Compared to the LMs fine-tuned using only the original training set (i.e., the Baseline column in Table~\ref{tab:rq3_fault_localization}), the LMs fine-tuned with both the training set and generated data demonstrate significant improvements in F1-score, Recall, Precision, Accuracy, and FPR.
For example, by adding 30\% of the generated data to the training set, GraphCodeBERT's F1-score increases from 0.384 to 0.550, Recall increases from 0.353 to 0.522, Precision increases from 0.421 to 0.582, and Accuracy increases from 0.889 to 0.917.

\textbf{Adding 30\%-50\% of generated data yields the best results for LMs.}
Generally, the performance of LMs improves with an increase in the amount of data, but it typically peaks when adding 30\%-50\% of generated data.
For example, when adding 10\% of generated data to the original training set, the Recall metric of CodeT5 improves by 1.91\% (i.e., 0.374 v.s. 0.367). 
When adding 30\% of data, the improvement is 47.14\% (i.e., 0.540 v.s. 0.367). 
However, when add 50\% and all of the generated data, the improvement diminishes (i.e., 0.496 v.s. 0.367 and 0.475 v.s. 0.367).
This discrepancy arises because the data is sorted, and the quality of the data decreases progressively. 
Therefore, too much data (i.e., All Generated) can introduce a significant amount of noise, while too little data  (i.e., 10\%×Generated) is insufficient for the LLMs to learn enough defect patterns.

As shown in Table~\ref{tab:rq3_additional}, \textbf{we find that LLM-generated data can enhance the fault localization capability of LMs on unknown real-world datasets.}
Similar to the results on Defects4J, adding a portion of the generated data yields the best performance. 
For example, adding 10\% of generated data to the original training set can increase the average F1-score of the LMs from 0.250 to 0.281.

\begin{table}[htbp]
  \centering
  \caption{Average increase (eight LMs) in fault localization on real-world datasets (RQ3)}
  \resizebox{.8\linewidth}{!}
  {
    \begin{tabular}{l|cccc}
    \toprule
    \textbf{Datasets} & \textbf{F1-score} & \textbf{Recall} & \textbf{Precision} & \textbf{Accuracy} \\
    \midrule
    Baseline & 0.250 & 0.196 & 0.361 & 0.852 \\
    10\%×Generated & \cellcolor[HTML]{C9E4B4}\textbf{0.281 (↑12.49\%)} & \cellcolor[HTML]{ADD88D}\textbf{0.235 (↑20.08\%)} & \cellcolor[HTML]{E3F2D9}\textbf{0.372 (↑3.20\%)} & 0.848 (↓0.45\%) \\
    30\%×Generated & \cellcolor[HTML]{C9E4B4}\textbf{0.278 (↑10.99\%)} & \cellcolor[HTML]{ADD88D}\textbf{0.243 (↑24.30\%)} & \cellcolor[HTML]{E3F2D9}\textbf{0.363 (↑0.57\%)} & 0.840 (↓1.37\%) \\
    50\%×Generated & \cellcolor[HTML]{E3F2D9}\textbf{0.263 (↑5.19\%)} & \cellcolor[HTML]{E3F2D9}\textbf{0.211 (↑7.82\%)} & \cellcolor[HTML]{E3F2D9}\textbf{0.371 (↑2.86\%)} & 0.851 (↓0.10\%) \\
    All Generated & \cellcolor[HTML]{E3F2D9}\textbf{0.254 (↑1.64\%)} & \cellcolor[HTML]{E3F2D9}\textbf{0.197 (↑0.47\%)} & \cellcolor[HTML]{E3F2D9}\textbf{0.369 (↑2.37\%)} & \cellcolor[HTML]{E3F2D9}\textbf{0.853 (↑0.17\%)} \\
    \bottomrule
    \end{tabular}%
  }
  \label{tab:rq3_additional}%
\end{table}%

\textbf{\underline{Improvement of Clone Detection.}}
As shown in Table~\ref{tab:rq3_clone_detection}, \textbf{we find that LLM-generated data can also improve the performance of clone detection.}
Specifically, after learning from the LLM-generated data, LMs show an improvement of 0.72\%$\sim$6.09\% on the POJ104 dataset. 
Similarly, on the Java250 dataset, LMs exhibit an increase of 0.12\%$\sim$3.23\%. 
Among all the LMs, CodeT5 and CodeT5+ perform the best.
For example, CodeT5+ achieves a MAP@R metric of 91.97\% on the POJ104 dataset, and CodeT5 achieves a MAP@R metric of 85.68\% on the Java250 dataset.
In total, the improvements for encoder-only LMs are significantly greater than those for encoder-decoder LMs. 
Encoder-only LMs achieve enhancements ranging from 0.44\% to 6.09\%, whereas encoder-decoder LMs only see improvements between 0.12\% and 2.21\%.
We believe this disparity may be attributed to the inherently superior performance of encoder-decoder LMs compared to encoder-only LMs.

\begin{table}[htbp]
  \centering
  \caption{Increase of MAP@R in clone detection (RQ3)}
  \resizebox{.7\linewidth}{!}
  {
    \begin{tabular}{l|cc|cc}
    \toprule
    \multirow{2.5}{*}{\textbf{Models}} & \multicolumn{2}{c|}{\textbf{CodeXGLUE-POJ104}} & \multicolumn{2}{c}{\textbf{CodeNet-Java250}} \\ 
    \cmidrule{2-5}
    & \textbf{Baseline} & \textbf{Generated} & \textbf{Baseline} & \textbf{Generated} \\
    \midrule
    BERT & 78.65\% & \cellcolor[HTML]{588E31}\textbf{83.44\% (↑6.09\%)} & 73.62\% & \cellcolor[HTML]{ADD88D}\textbf{75.19\% (↑2.13\%)} \\
    RoBERTa & 81.65\% & \cellcolor[HTML]{588E31}\textbf{84.18\% (↑3.10\%)} & 77.13\% & \cellcolor[HTML]{588E31}\textbf{79.62\% (↑3.23\%)} \\
    CodeBERT & 84.93\% & \cellcolor[HTML]{ADD88D}\textbf{87.24\% (↑2.72\%)} & 80.98\% & \cellcolor[HTML]{C9E4B4}\textbf{81.89\% (↑1.12\%)} \\
    GraphCodeBERT & 83.55\% & \cellcolor[HTML]{588E31}\textbf{87.37\% (↑4.57\%)} & 83.09\% & \cellcolor[HTML]{E3F2D9}\textbf{83.46\% (↑0.44\%)} \\
    UniXcoder & 90.50\% & \cellcolor[HTML]{C9E4B4}\textbf{91.76\% (↑1.39\%)} & 83.45\% & \cellcolor[HTML]{ADD88D}\textbf{85.37\% (↑2.30\%)} \\
    PLBART & 85.34\% & \cellcolor[HTML]{ADD88D}\textbf{87.11\% (↑2.07\%)} & 82.79\% & \cellcolor[HTML]{ADD88D}\textbf{84.62\% (↑2.21\%)} \\
    CodeT5 & 90.46\% & \cellcolor[HTML]{E3F2D9}\textbf{91.11\% (↑0.72\%)} & 85.58\% & \cellcolor[HTML]{E3F2D9}\textbf{85.68\% (↑0.12\%)} \\
    CodeT5+ & 90.52\% & \cellcolor[HTML]{C9E4B4}\textbf{91.97\% (↑1.60\%)} & 82.35\% & \cellcolor[HTML]{C9E4B4}\textbf{83.88\% (↑1.86\%)} \\
    \bottomrule
    \end{tabular}
  }
  \label{tab:rq3_clone_detection}%
\end{table}%

\intuition{
\textbf{Finding 3}: 
Incorporating LLM-generated data into the original training set leads to substantial performance improvements for LMs across both fault localization and code clone detection tasks.
}

\section{Discussion}
\label{sec:discussion}

\subsection{Data Selection Analysis}

\begin{figure*}[t]
    \centering
    \includegraphics[width=\linewidth]{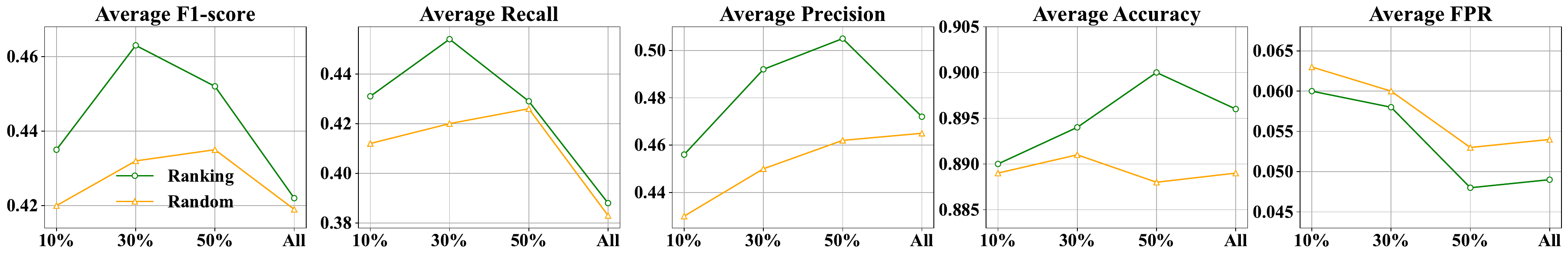}
    \caption{Average result (eight LMs) in fault localization when using ranking or random selection}    \label{fig:dicussion_fault_localization}
\end{figure*}

We utilize LLMs to generate a significant amount of candidate data for fault and clone generation.
However, the quality and distribution of this data vary, so we have devised selection strategies to choose high-quality data.
In this section, we discuss the role of these selection strategies.
For fault localization, we compare our designed ranking selection strategy with random selection. 
In the random selection strategy, for each non-buggy function, we randomly select one buggy function from the corresponding buggy functions to pair with it and randomly choose 10\%, 30\%, and 50\% of the data as needed.
For clone detection, we compare the LMs fine-tuned on the selected generated data with those fine-tuned on the original generated data.
Fig.~\ref{fig:dicussion_fault_localization} presents the average results (eight LMs) in fault localization when using ranking or random selection. 
We see that compared to random selection, ranking selection shows better performance across five metrics. 
This demonstrates that our designed score (refer to Section~\ref{sec:approach} for more detail) can be an effective measure to rank potentially high-quality data.
Table~\ref{tab:dicussion_clone_detection} shows the average results (eight LMs) in clone detection when adding all generated data or selected data. 
On the whole, the selected data performs better than the unselected data, improving by 0.78\%$\sim$2.71\% compared to the baseline.
Overall, our selection strategies showcase their efficacy in enhancing both fault localization and clone detection tasks, underscoring the importance of thoughtful data processing in maximizing model performance.

\begin{table}[htbp]
  \centering
  \caption{{Average result (eight LMs) in clone detection when adding all generated data or selection data}}
  \resizebox{.6\linewidth}{!}
  {
    \begin{tabular}{lcc}
    \toprule
    \textbf{Datasets} & \textbf{CodeXGLUE-POJ104} & \textbf{CodeNet-Java250}  \\
    \midrule
    Baseline & 85.70\% & 81.82\% \\
    Generated (All) & \cellcolor[HTML]{C9E4B4}\textbf{86.89\% (↑1.39\%)} & \cellcolor[HTML]{E3F2D9}\textbf{82.15\% (↑0.40\%)} \\
    Generated (Selection) & \cellcolor[HTML]{ADD88D}\textbf{88.02\% (↑2.71\%)} & \cellcolor[HTML]{E3F2D9}\textbf{82.46\% (↑0.78\%)} \\
    \midrule
    CloneGen (Overlap) & \cellcolor[HTML]{E3F2D9}\textbf{{86.18\% (↑0.56\%)}} & - \\
    Generated (Overlap) & \cellcolor[HTML]{C9E4B4}\textbf{{86.99\% (↑1.51\%)}} & - \\
    \bottomrule
    \end{tabular}%
  }
  \label{tab:dicussion_clone_detection}%
\end{table}%

\subsection{Compare with Traditional Approaches}

Traditional generation tools can also be used to generate data. 
In this discussion, we aim to compare the quality of the data generated by LLMs with that of data generated by generation tools.

For fault generation, we use the Major~\cite{just2014major} 
mutation tool to generate faults. 
Major is an efficient and flexible mutation analysis framework that supports generating mutants during compilation and exporting source-code mutants. 
Major supports eight mutation operators: AOR, LOR, SOR, COR, ROR, ORU, LVR, and STD. 
We perform mutation on all fixed versions of the programs in Defects4J and then extract the mutants (i.e., the buggy functions) that are killed by the test cases.
We use two methods to select buggy functions from the mutated results. 
The first method employs our proposed selection strategy to choose the optimal data, while the second method involves randomly selecting one buggy function from all mutated candidate buggy functions to form a data pair with the corresponding non-buggy function.
To ensure a fair comparison,
we select the intersection of LLM-generated data and Major-generated data, retaining only the pairs of non-buggy and buggy functions where the non-buggy function appears in both data.
This process resulted in 3,619 pairs of non-buggy and buggy functions. 
From the results in Table~\ref{tab:disussion_2}, we find that LLM-generated data generally outperforms mutated data. 
Although mutated data is better at improving the Precision of LMs, it leads to a significant drop in F1-score and Recall.
This might be due to two reasons: 
(1) all mutation operations are implemented on a single line, whereas LLMs can insert faults across multiple lines; 
(2) the faults generated by mutation follow patterns, resulting in less diverse defects.

For clone generation, we use the CloneGen~\cite{zhang2023challenging} approach to generate code clones. 
CloneGen applies 15 atomic transformation operators to transform the C/C++ code, generating clones while preserving the original code's semantics.
We transform the ground truth for each task in HumanEval-X-C++ and perform multiple transformations; for instance, a clone code generated by the first transformation operator can be further transformed by a second operator to create additional clones.
To ensure a fair comparison, we ensure that each task had the same number of clones in both CloneGen-generated data and LLM-generated data.
Table~\ref{tab:dicussion_clone_detection} shows the performance improvement of LMs using clones generated by the CloneGen method and those generated by LLMs. 
We find that with the same number of clones, LLM-generated data has a better effect, improving LM performance by an average of 1.51\%. 
In contrast, CloneGen-generated data only improved performance by 0.56\%.

\begin{table}[htbp]
  \centering
  \caption{{Average result (eight LMs) in fault localization when comparing with mutation tool}}
  \resizebox{.8\linewidth}{!}
  {
    \begin{threeparttable} 
    \begin{tabular}{l|cccc}
    \toprule
    \textbf{Datasets} & \textbf{F1-score} & \textbf{Recall} & \textbf{Precision} & \textbf{Accuracy} \\
    \midrule
    Baseline & 0.392 & 0.384 & 0.406 & 0.882 \\
    Mutated & 0.366 (↓6.63\%) & 0.276 (↓28.13\%) & \cellcolor[HTML]{ADD88D}\textbf{0.555 (↑36.70\%)} & \cellcolor[HTML]{C9E4B4}\textbf{0.906 (↑2.72\%)} \\
    Mutated* & 0.360 (↓8.16\%) & 0.291 (↓24.22\%) & \cellcolor[HTML]{ADD88D}\textbf{0.489 (↑20.44\%)} & \cellcolor[HTML]{E3F2D9}\textbf{0.896 (↑1.59\%)} \\
    Generated & \cellcolor[HTML]{ADD88D}\textbf{0.438 (↑11.73\%)} & \cellcolor[HTML]{E3F2D9}\textbf{0.391 (↑1.82\%)} & \cellcolor[HTML]{ADD88D}\textbf{0.508 (↑25.12\%)} & \cellcolor[HTML]{C9E4B4}\textbf{0.902 (↑2.27\%)} \\
    \bottomrule
    \end{tabular}%
    ``Mutated*'': Randomly select from Major-generated data.
    \end{threeparttable}
  }
  \label{tab:disussion_2}%
\end{table}%

\subsection{Threats to Validity}

\noindent
\textbf{Internal Validity.}
The first one is the design of the prompt to instruct LLMs to generate responses.
Our prompt design is based on practical advice~\cite{shieh2023best,nijkamp2022codegen, li2023starcoder,roziere2023code,yin2024multitask}, which has been validated by numerous users online and has shown to elicit good responses from LLMs.
The second one is about the potential mistakes in the implementation of studied LMs and LLMs. 
To mitigate such threats, we use the original source code provided by the corresponding authors, ensuring a direct and reliable foundation for our analysis.

\noindent
\textbf{External Validity.}
One potential threat to external validity is the generalizability of our findings across different datasets, programming languages, and model architectures.
Our study involves datasets for both Java and C++, which are widely used in previous research on fault localization and clone detection~\cite{yang2024large,guo2022unixcoder,wang2023codet5+}. 
We utilize a diverse array of pre-trained LMs, including both encoder-only and encoder-decoder architectures, with variations in their parameter sizes and classification capabilities. 
Moreover, our study includes eight distinct LLMs, comprising both code LLMs and general LLMs, each offering unique attributes due to differences in their parameter sizes and generation capabilities.
Despite the breadth of our experimentation, it remains uncertain whether our findings can be broadly generalized to other contexts. 
Thus, further investigation is necessary to validate and extend the conclusions drawn from our research.

\section{Conclusion}
\label{sec:conclusion}

In this paper, we leverage the powerful generation capabilities of LLMs to enhance pre-trained LMs. 
Specifically, we employ LLMs to produce domain-specific data, thereby bolstering the efficacy of pre-trained LMs on the target tasks.
We conduct experiments by combining different LLMs in our generation phase and introducing various LMs to learn from the LLM-generated data. 
Then, we compare the performance of these LMs before and after learning the data.
We find that LLM-generated data significantly enhances the performance of LMs. 
The improvement can reach up to 58.36\% for fault localization and up to 6.09\% for clone detection.
This investigation underscores the considerable potential of leveraging LLMs for data generation to yield significant performance gains in LMs.

\balance
\bibliographystyle{ACM-Reference-Format}
\bibliography{main}

\end{document}